# What happens to psychological safety when going remote?

Anastasiia Tkalich, SINTEF, Trondheim, 7034, Norway

Darja Smite, Blekinge Institute of Technology, 371 41, Sweden

Nina Haugland Andersen, Norwegian University of Science and Technology, 7034, Norway

Nils Brede Moe, SINTEF, Trondheim, 7034, Norway

*Psychological safety is a precondition for learning and success in software teams. Companies such as SavingsBank, which is discussed in this article, have developed good practices to facilitate psychological safety, most of which depend on face-to-face interaction. However, what happens to psychological safety when working remotely? In this article, we explore how Norwegian software developers experienced pandemic and post-pandemic remote work and describe simple behaviors and attitudes related to psychological safety. We pay special attention to the hybrid work mode, in which team members alternate days in the office with days working from home. Our key takeaway is that spontaneous interaction in the office facilitates psychological safety, while remote work increases the thresholds for both spontaneous interaction and psychological safety. We recommend that software teams synchronize their office presence to increase chances for spontaneous interaction in the office while benefitting from focused work while at home.*

Professional work life has likely split irreversibly into pre-pandemic and post-pandemic times, as the better-than-expected work from home (WFH) pandemic experiences[1] changed the prevalence and perception of remote work in software companies. Society's reopening clearly demonstrates that employees' willingness to return to the office differs greatly. Some return to their pre-pandemic routines, some visit the office only a few days per week and spend most of their time working from home, and others prefer to continue to WFH indefinitely. As a result, predominantly onsite work routines are being adjusted to the "new norm" or the hybrid work mode.

Remote work is not a new phenomenon. The first studies on "teleworking" appeared in the 70s and regarded it as a temporary and often partial practice chosen by few[2]. The rise of popularity of offshoring in the late 90s increased the prevalence of remote work as projects became distributed[3]. Yet, remote WFH has never been so widespread. Thus, questions related to the impact of WFH on software development practices have been in the spotlight of research and practice. So far, WFH research has focused on individual experiences, primarily productivity[4,5,6] and well-being[7], identifying more benefits than challenges. However, individual gains in productivity enabled by the absence of office interruptions have been debated in team-oriented research that highlights the importance of constant connectivity among teammates[8]. Teams in fully remote WFH mode suffer from a limited ability to brainstorm, difficulty communicating, and decreased satisfaction with interactions from social activities[9]. Similarly, prior work on partially dispersed teams (relevant in the hybrid work mode, in which developers alternate their office days with WFH) suggests that remoters tend to have significantly reduced team cohesion, poor awareness of "who did what" and "who knows what," divergent viewpoints, conflicts, and coordination problems[10]. While the hybrid workplace promises to reduce remote working due to episodical co-located work, there is no clear understanding of the impact of hybrid work on successful team functioning.

Psychological safety (see the Sidebar) is among the social factors that might be threatened by remote and hybrid work. While psychological safety contributes to team learning, commitment, and performance[11,12], little is known about how remote and hybrid work affect team psychological safety. Given that the hybrid workplace is our future, this knowledge might be crucial for avoiding painful mistakes.

In this article, we describe how psychological safety is affected by hybrid work in SavingsBank (company's name is anonymized for confidentiality reasons). Lessons learned from our study comprise organizational practices that increase psychological safety in teams working onsite and remotely as well as changes in team behaviors and attitudes in different work modes.

## Sidebar: What is psychological safety?

Psychological safety is an important cognitive and interpersonal concept that relates to several positive outcomes in software teams. Psychological safety is defined as "a shared belief held by members of a team that the team is safe for interpersonal risk-taking."[13] A psychologically safe environment minimizes the potential negative consequences of making mistakes or taking initiative, which refocuses teams on a task instead of interpersonal problems and thus improves performance. Psychological safety is positively related to information sharing, learning behaviors, employee engagement, satisfaction, and commitment of team members[11]. Teams with high psychological safety perform better than those in which it is low[12].

Edmondson made a fundamental contribution to defining and measuring team psychological safety in 1999[13], explaining it briefly as a sense of confidence among team members that fellow teammates will not embarrass, reject, or punish them for speaking up. She suggested seven items combining behavioral, affective, and cognitive aspects, including "*It is safe to take a risk on this team*" and "*It is difficult to ask other members of this team for help.*" Inspired by these items and Edmondson's work, we formulated four core dimensions that we believe collectively reflect how psychological safety can be observed in the behaviors and attitudes of members of software teams (read more in the supplemental material). The core dimensions are as follows:

1. **Safe to be honest:** Teammates share ideas, opinions, and concerns and bring up problems and tough issues without fear of social penalty.

2. **Safe to make mistakes:** Teammates perceive it to be okay to make mistakes and do not blame but instead focus on learning from constructive feedback.

3. **Safe to ask for help:** Teammates perceive that it is easy to ask others for help.

4. **Valuing each other:** Teammates believe that their efforts will not be deliberately undermined (downgraded, overlooked, or ignored), have a positive attitude toward each other's contributions, give frequent positive feedback, and acknowledge each member's skills, talents, and inputs.

## The SavingsBank Case

We performed a longitudinal study of three teams from SavingsBank during and after the COVID-19 pandemic (fall 2020-winter 2022). SavingsBank is a software development company owned by an alliance of banks that employed 24 software teams in two locations in Norway. In March 2020, employees moved from predominantly onsite work to 100% WFH due to government-enforced regulations. Offices were open only for special needs (for example, to run tests on a specific network). For all, offices reopened and closed repeatedly in response to the reduced or increased virus spread (open in the summer 2020 and 2021, closed in the fall 2020 and 2021). When open, offices had certain restrictions (limited presence in the offices and in meeting rooms; social distancing) and rather low occupation, not exceeding 60% (see Figure 1). In February 2022, the offices opened without restrictions. Since then, teams are expected at the office at least two days per week, which the teams can choose themselves.

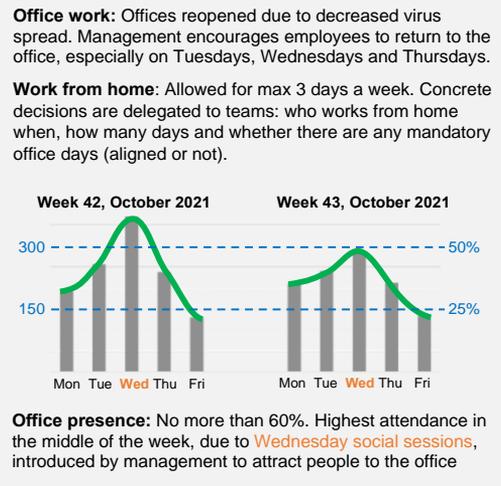

**Office work:** Offices reopened due to decreased virus spread. Management encourages employees to return to the office, especially on Tuesdays, Wednesdays and Thursdays.

**Work from home**: Allowed for max 3 days a week. Concrete decisions are delegated to teams: who works from home when, how many days and whether there are any mandatory office days (aligned or not).

**Office presence:** No more than 60%. Highest attendance in the middle of the week, due to Wednesday social sessions, introduced by management to attract people to the office

FIGURE 1. Employee attendance in fall 2021

Due to work being distributed across two Norwegian sites pre-pandemic, SavingsBank had the appropriate conditions and infrastructure to move to WFH. However, remote work had not been common in all teams, and not everyone had established practices for collaboration outside of the office walls. SavingsBank's focus on autonomous working was another facilitator of WFH. Practitioners describe SavingsBank as a leading agile environment in Norway using state-of-the-art collaboration methods and technologies and focusing on teams' autonomy, as evident in the following quote:

> *We made some general rules, but were cautious with telling teams how to organize their work*

*[onsite vs. WFH]. Teams are allowed to find what suits them best. [Development manager]*

We conducted 16 semi-structured interviews with various stakeholders: six interviews with members of three software teams (in-house developers, a designer, and consultants) in February-March 2021 (remote mode), six interviews in November 2021-January 2022 (hybrid mode), and four interviews with the HR manager and the development manager (twice each). Participants' team experience ranged between several months and 2.5 years. We also collected the leadership group's feedback on our findings and analyzed secondary survey data gathered by SavingsBank between summer 2020 and summer 2022 (developers' preferences for WFH, job satisfaction, meetings data, office attendance, and strategy documents) and other internal documents. For more details, see the supplementary material.

## SavingsBank practices that support psychological safety

Psychological safety was a strategic focus area for SavingsBank, even pre-pandemic. Herein, we describe corporate routines facilitating psychological safety, which were found useful especially when working remotely.

**Pull requests** (PRs), the standard mechanism for reviewing code, facilitate the exchange of feedback on code changes. Reviewers and contributors' interactions are documented in discussion threads and are available to all. PRs foster knowledge sharing, early feedback, and increased collaboration and reduce the risk of putting flawed code into production. Developers describe PRs as a means of sharing responsibility if something goes wrong:

*PR is a good safety net […] when I write code, there is nothing that is put into production unless others have looked at it too. [Developer, Team C]*

**Monday commitments and FridayWins** are weekly team meetings. Monday commitments gather teammates for ~15 minutes to discuss progress toward team goals and weekly focus to reach the goal, and they prioritize tasks for inclusion in the board. FridayWins (~30 minutes) focus on celebrating teams' achievements and are important for increasing psychological safety:

*The whole point of FridayWins is to acknowledge each other´s achievements. [Developer, Team B]*

**Postmortems** are team meetings that follow a significant failure. Several participants described postmortems as crucial for an environment in which it is *safe to make mistakes* because it helps avoid scapegoating and strengthens organizational norms of learning from mistakes:

*Postmortems are really great; they do not only emphasize that we should not blame each other, but also [help us] achieve it in practice. [Developer, Team A]*

**20% time** has been SavingsBank's policy since 2018, allowing employees to dedicate one weekday to building new competencies and developing architecture, tools, and practices together. Participants described it as positive in terms of feeling professionally valued by the company:

*It is unique that we can use one weekday like this; I feel my company appreciates my skills and supports further professional development. [Developer, Team B]*

**Demos with customers** allow teams to regularly present and discuss the latest features and facilitate professional acknowledgement by colleagues:

*We often get comments after such demos, for example, on Slack, where our team lead can say "Great job today, Tommy," and then others join and add clapping and thumbs up icons, as thanks. [Developer, Team A]*

## Team behaviors when going hybrid

We found that hybrid modes were twofold: misaligned mode, in which members decided when to work in the office or from home impromptu, and aligned mode, in which members had agreed-upon common office days. In the following, we describe how teammates used organizational practices and reflected on the changes in behaviors and attitudes along four dimensions of psychological safety, outlined in the Sidebar, in different work modes: onsite, remote, or hybrid (See Figure 2).

| SBP* | Behaviours and attitudes | Onsite – Work in the office | Hybrid – A mix of remote work and work in the office | Remote – Work from home, café, co-working space, etc. |
|---|---|---|---|---|
| Safe to be honest | Speaking up in meetings | Low threshold for engaging in a conversation and expressing own opinion | Unequal experiences: remoters are likely to remain in the listening mode. | Increased threshold for engaging in an online conversation |
| Safe to be honest | Providing critical feedback or disagreeing | Harder to criticize face-to-face (F2F), but oral feedback is perceived less negatively than the written feedback. | Unequal experiences: remoters are likely to not receive feedback on their actions. | Chances of withholding critical feedback or providing too much critical feedback. |
| Safe to be honest | Asking "stupid questions" | Low threshold to ask all sorts of questions F2F | Unequal experiences: remoters are likely to withhold or delay their questions until they can be asked F2F. | Increased threshold for asking "stupid questions" in group channels due to public exposure; and longer time for formulating questions. |
| Safe to make mistakes (Pull requests (PRs), postmortems) | Requesting feedback on unfinished work results | Low threshold for showing unfinished work due to the ease to reach out and better chances to follow up with explanations and discussions (also by the whiteboard). | Unequal experiences: remoters are likely to delay discussing unfinished work until in the office, and/or proceed with the work without intermediate feedback. | Decreased chances to share unfinished because of a higher threshold for reaching out |
| Safe to make mistakes | Not blaming peers for making mistakes | The established culture of shared responsibility for the quality of the work outcomes is strong and does not depend on the working mode. Neither remote members, nor onsite members are blamed for making mistakes. Mistakes are discussed in code reviews (in writing) or in post-mortems (orally). The latter are held on site, remotely or as a hybrid meeting. However, the mode of the meeting does not affect the peers' attitudes or the way mistakes are discussed. Mistakes are treated as opportunities to learn. | | |
| Safe to ask for help | Asking each other for help | Low threshold for asking for help F2F due to the availability of contextual clues about people being busy, and lower thresholds and increased chances for initiating interaction. Help requests are handled promptly, written help requests are easily followed up in-person in case of a delay. | Unequal experiences: remoters are unlikely to be spontaneously asked for help, or contribute to problem-solving. Requests for help from remoters are usually delayed until F2F interaction or scheduled online meetings. | Increased threshold for asking for help: chats are perceived inefficient while video/phone calls have a higher threshold for initiation due to a lack of visibility into one's availability (especially for newly employed). Help is often delayed, thus peers are unlikely to ask or receive help spontaneously. |
| Valuing each other (Friday Wins, demos with customers, 20%time, PRs) | Giving each other positive feedback and acknowledging effort | F2F interactions strengthen the emotional effect of positive feedback, and the likelihood of team members willing to compliment each other and thank for the help provided. | Unequal experiences: remoters are likely to receive fewer compliments about their work results. | Decreased chances to receive positive feedback, and lower emotional effects, both when shared in the writing (emojis or comments) and orally in a video meeting. |
| Valuing each other | Feeling of belonging in the team | Stronger feeling of belonging and more chances to build informal bonds F2F, as well as the chances for joint gatherings and willingness to be inclusive. | Unequal experiences: remoters are likely to be alienated and excluded from the spontaneous discussions and celebrations in the office. Remote work is justly associated with the fear of missing out (FOMO). | Decreased feeling of belonging, especially among the new hires, and increased difficulty to have spontaneous gatherings. |

*SavingsBank practices that support psychological safety

FIGURE 2. Behaviors and attitudes that indicate psychological safety in various work modes

#1. Safe to be honest

**Speaking up in meetings** was perceived to be easier face-to-face than remotely. We learned that the absence of non-verbal cues makes some more hesitant to participate in discussions and more focused on listening, while technical delays in communication make people hold back due to the fear of interrupting. Online meetings demand more structure and effort to stimulate engagement, while onsite meetings facilitate a more dynamic and spontaneous exchange of opinions:

> *Clearly, the threshold for joining a discussion is much higher when you join a digital meeting compared to when you join physically. [...] If several people are participating physically, you can easily notice that the digital group becomes a little passive because the activity in the room overruns what digital participants are saying. [Designer, Team A]*

**Providing critical feedback.** When working remotely, team members relied heavily on written communication, which created barriers to providing critical feedback because of frequent misunderstandings:

> *The written format makes things ruder and more negative than meant. If you criticize in a physical mode, one can easily solve it by discussing. [Designer, Team A]*

To avoid misunderstandings, many preferred to follow written feedback up with a conversation, which is easier to accommodate in the office. In the hybrid mode, we learned that remoters are unlikely to share critical feedback, and if sharing, rarely follow it up because of the hurdles of scheduling online meetings. In fact, the lack of spontaneous interaction in remote and hybrid modes was mentioned frequently as the major impediment. Evidently, in the hybrid mode, remoters are likely to work in isolation without frequent feedback.

**Asking "stupid" questions.** Several informants mentioned the importance of being able to ask silly questions as a sign of psychological safety in a team. Interestingly, we found that this was never easy due to the possible embarrassment. In the office, silly questions can be overheard, while questions posted in team channels can be read by everyone, leading to possible public shame. This was especially problematic for new hires and inexperienced members who feared revealing their lack of competence and had higher barriers to approach others. We learned that asking silly questions required training and was easier in personal inquiry (e.g., by using chats and scheduling short video calls):

> *Now I tend to just send a message and ask if one has time for a small conversation. [...] I started to ask "stupid" questions as early as possible just to be less afraid. [Developer, Team A]*

#### #2. Safe to make mistakes

**Requesting feedback on unfinished work.** When team members show unfinished design sketch or code, they are likely to expose mistakes. In the SavingsBank teams, members were not afraid to reveal such mistakes. We also learned that feedback requests are often spontaneous and thus easier to make in the office, such as by inviting someone to look at a screen or whiteboard. Sharing unfinished work remotely was said to be harder because it takes additional effort to initiate and because of the possible delay in receiving feedback:

> *If one shows a design at the office, one can receive instant feedback and maybe explain why one did what one did. This is way easier than to write a whole thesis on Slack. [Designer, Team A]*

**Not punishing or blaming team members for making mistakes.** All our participants agreed that they did not fear making mistakes because of well-established routines and the culture of treating mistakes as learning opportunities, and discussing them openly in dedicated postmortems. Further, teams cultivated shared responsibility for code quality, fostered by code review routines (PRs):

> *If something is wrong that is approved and goes all the way out, then [...] no one becomes a scapegoat, and you all solve it together. [Developer, Team B]*

#### #3. Safe to ask for help

**Asking each other for help.** Any request directed to teammates leads to disturbances, and we found that disturbing others in the office was easier due to contextual clues that help evaluate whether people are busy or not. Such clues were absent when working remotely or for remoters in the hybrid mode, for whom the threshold for asking for help increases. The absence of spontaneous peer help can lead to additional time spent on solving problems in isolation. Interestingly, we found that experiences of onsite or hybrid teamwork among the new hires increased their chances of exhibiting psychologically safe behaviors when working remotely. A newly hired developer explained,

> *It takes me longer to ask for help when I sit at home. [...] But I felt safer in asking for help digitally because I have now spent some time with the team physically and had some informal talks. [Developer, Team B]*

Further, a worker´s likelihood of asking for help depended on the perceived timeliness of the feedback. In the office, help requests were processed within seconds or easily followed by a reminder (knocks on the door, comments at the coffee machine). When working remotely (in remote and hybrid modes), members could not be certain whether and when the feedback would come. A senior developer explained that response times and ways to handle requests varied greatly:

> Some are like "I am a little busy now, but I will look at it ASAP." Others you don´t get much response from, and in a couple of days, they return with a perfect answer with tons of details. [...] Some become bottlenecks, I wait and nag... [Developer, Team A]

#### #4. Valuing each other

**Giving each other positive feedback and acknowledging the effort.** Positive feedback was important to feel valued in the team. Sharing positive feedback face-to-face was found to be effortless, and the emotional effect of such feedback was stronger than that of written feedback shared remotely or in online meetings. In fact, the probability of receiving extensive positive feedback remotely decreased. We found that positive feedback in team channels was often limited to the use of digital icons:

> ... then many thumbs up and celebrative emojis appear. So, there are many such micro-interactions. [Developer, Team C]

PR comments were another way to acknowledge contributions:

> I can answer the PR comments like, "Wow, I haven´t thought about it!" and sometimes maybe I say "Thanks." [Developer, Team B]

**Feeling of belonging in the team.** Participants agreed that the feeling of being valued contributed to the feeling of belonging, which was strengthened through numerous spontaneous office interactions and exchange of positive feedback. When working remotely, the feeling of belonging was weaker, especially among new hires who had not managed to develop informal bonds:

> In the office, you get to talk about what you did on the weekend and what goes on in life. This leads to stronger bonds and to the feeling of belonging. [Designer, Team A]

We also found that the hybrid mode can alienate remoters, as they are excluded from spontaneous office discussions and have a feeling of missing out on what is going on in the team.

## What have we learned?

Perhaps not surprisingly, we found remote work to hinder psychological safety in teams. Overall, we conclude that many behaviors and attitudes related to psychological safety are fostered by spontaneity of interaction (see Figure 3). Office presence facilitates spontaneous questions, requesting/giving feedback, speaking up, ideating, and including each other. The threshold for such behaviors is higher when working remotely, as digital interactions require upfront planning, waiting for a response, arranging the digital setup, or overcoming technical issues. Besides, the absence of contextual clues for remoters makes it harder to time interactions to avoid unwelcome interruptions. Interestingly, the hybrid mode increases the chances of spontaneous interaction for those in the office, but simultaneously leads to splitting the team into subgroups or alienating individual members, thus reducing the team's psychological safety. However, do not get too pessimistic; we also learned that teams with synchronous office presence experience fewer challenges, are able to interact spontaneously when working onsite, and tap into the benefits of remote WFH without alienation or division into subgroups.

For those concerned about the practicability of hybrid work, we would like to emphasize a few important benefits of WFH, when done wisely. Some interviewees associated WFH with increased individual productivity thanks to the ability to stay focused, as found in related research[6]. Interruptions, which are common when working face-to-face, may cost up to 10-15 min in terms of regaining one's focus[14], and one rarely has control over how often and when they are interrupted. Although remote work can be challenging for some, we found that teams adjust their tactical approaches to overcome these challenges successfully. Our participants used break-out rooms to foster active participation, implemented Huddle to contact each other more easily, and made many other adjustments for convenient remote work.

For teams, we recommend finding effective ways to promote the onsite collaboration and spontaneous interactions necessary for psychological safety alternated with vital time for focused individual work remotely. However, we have no golden rule for how many days in the office/home are optimal. We suggest focusing future research on finding ways to simulate spontaneity of computer-mediated interaction, for example, by adapting tools that provide additional contextual clues (open video rooms, status labels[15]).

Finally, we found that organizational norms and routines are effective at supporting team psychological safety. Institutionalized practices such as post-mortems and PRs support the safety to make mistakes, while FridayWins, 20% time, and customer demos help members feel valued. We also found that such practices are equally powerful when working onsite and remotely. Other practices that help with psychological safety are mandatory office days and forming teams with synched preferences for office presence, which are important directions for further study. For now, we found one successful case in SavingsBank, when a developer switched teams to work with like-minded teammates. We believe such actions can help companies balance the need to maintain strong teams and provide flexibility in choosing effective individual work modes

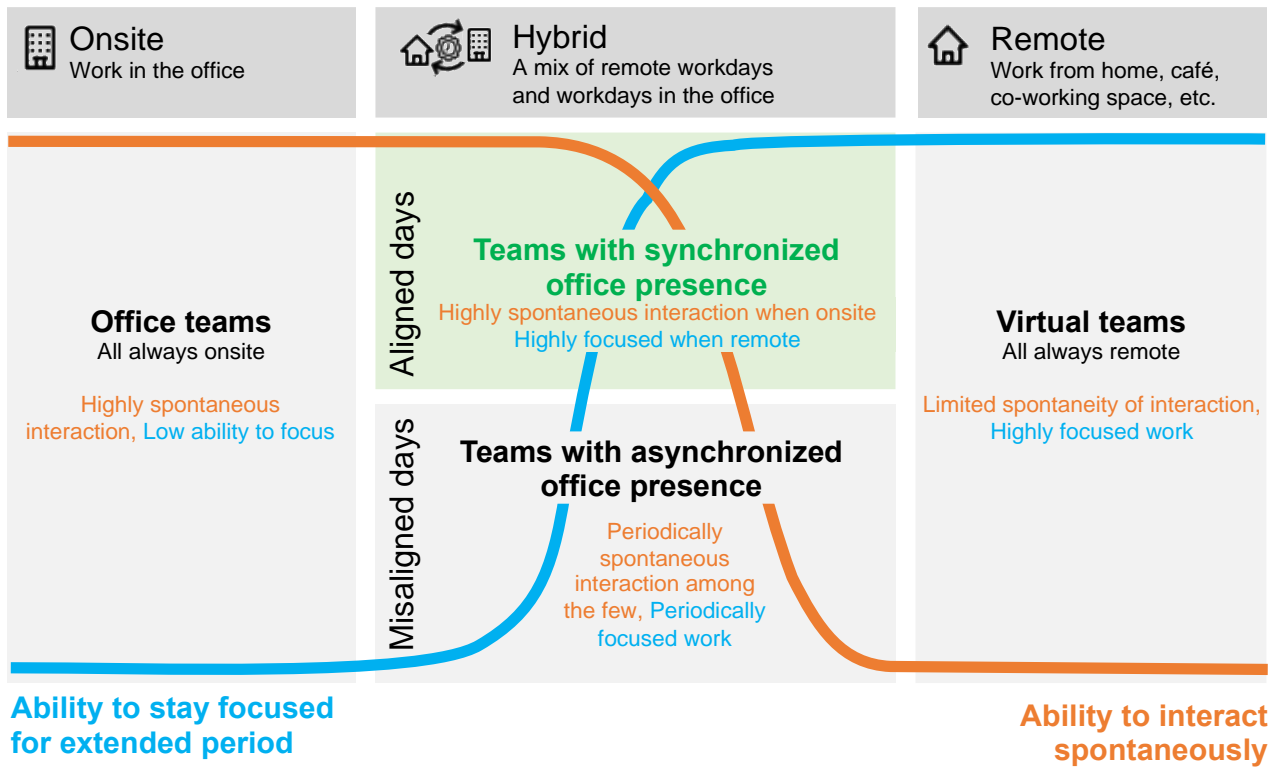

FIGURE 3. A relationship between focus and ability for spontaneous interaction in different work modes